\documentclass[sigconf, nonacm]{acmart}

\newcommand\vldbdoi{XX.XX/XXX.XX}
\newcommand\vldbpages{XXX-XXX}
\newcommand\vldbvolume{18}
\newcommand\vldbissue{1}
\newcommand\vldbyear{2025}
\newcommand\vldbauthors{\authors}
\newcommand\vldbtitle{\shorttitle} 
\newcommand\vldbavailabilityurl{}
\newcommand\vldbpagestyle{plain} 

\usepackage{enumitem}
\usepackage{listings}
\usepackage{xcolor}
\usepackage{xspace}
\usepackage{multirow}
\usepackage{makecell}
\usepackage{adjustbox}
\usepackage{tikz}

\newif\ifshowcomment

\newcommand\calcite{Calcite\xspace}
\newcommand\cockroach{CockroachDB\xspace}
\newcommand\optgen{Optgen\xspace}
\newcommand\cosette{Cosette\xspace}
\newcommand\datafusion{DataFusion\xspace}

\newcommand\hottsql{HoTTSQL\xspace}
\newcommand\java{Java\xspace}

\newcommand\qedp{QED\xspace}
\newcommand\rulescript{\textsc{RuleScript}\xspace}
\newcommand\smt{SMT\xspace}
\newcommand\snowflake{Snowflake\xspace}
\newcommand\spes{SPES\xspace}
\newcommand\sql{SQL\xspace}

\newcommand\systemR{System R\xspace}

\newcommand\op[1]{\textsf{\upshape #1}}

\lstdefinestyle{mystyle}{
    backgroundcolor=\color{white},   
    commentstyle=\color{gray},
    keywordstyle=\color{violet},
    numberstyle=\numberstyle,
    stringstyle=\color{teal},
    basicstyle=\footnotesize\ttfamily,
    breakatwhitespace=false,
    breaklines=true,
    captionpos=b,
    keepspaces=true,
    numbers=left,
    numbersep=0em,
    showspaces=false,
    showstringspaces=false,
    showtabs=false,
}

\newcommand\numberstyle[1]{%
    \footnotesize
    \color{lightgray}%
    \ttfamily
    \ifnum#1<10 0\fi#1 |%
}

\lstdefinelanguage{erlqo}{
    morekeywords={RuleFamily, RuleGenerator, Rule, Plan, Empty, Values, Filter, Project, Join, Union, Aggregate, Distinct, Alias, PatnAlias, FOPredAlias, PredAlias, CallAlias, JtyAlias, Id, Field, Stat, Op, FORALL, EXISTS, NOT, AND, OR, IS, NULL, TRUE, FALSE},
    sensitive=false, % keywords are not case-sensitive
    morecomment=[l]{//}, % l is for line comment
    morecomment=[s]{/*}{*/}, % s is for start and end delimiter
    morestring=[b]", % defines that strings are enclosed in double quotes
    morestring=[s]{<}{>} % special keywords
} %

\begin{document}
%% Variable control
% Toggle whether to show a comment
\showcommenttrue
%\showcommentfalse

% Set code style
\lstset{language=erlqo, style=mystyle, upquote=true}

% Set list appearance
\setlist[itemize]{leftmargin=*, topsep=0.5em}
\setlist[enumerate]{leftmargin=*, topsep=0.5em}

% Set equation space
\setlength{\abovedisplayskip}{0.5em}
\setlength{\belowdisplayskip}{0.5em}

% Start of document
\title{An Extensible and Verifiable Language for Query Rewrite Rules}

%%
%% The "author" command and its associated commands are used to define the authors and their affiliations.
\author{Sicheng Pan}
\affiliation{\institution{University of California, Berkeley}\country{}}
\email{pansicheng@berkeley.edu}

\author{Shuxian Wang}
\affiliation{\institution{University of California, Berkeley}\country{}}
\email{wsx@berkeley.edu}
\author{Wesley Zheng}

\affiliation{\institution{University of California, Berkeley}\country{}}
\email{wzheng0302@berkeley.edu}

\author{Zirong Zeng}
\affiliation{\institution{The Chinese University of Hong Kong, Shenzhen}\country{}}
\email{zirongzeng@link.cuhk.edu.cn}

\author{Vijay Sharma}
\affiliation{\institution{University of California, Berkeley}\country{}}
\email{vijaysharma@berkeley.edu}

\author{Alvin Cheung}
\affiliation{\institution{University of California, Berkeley}\country{}}
\email{akcheung@cs.berkeley.edu}

%%
%% The abstract is a summary of the work to be presented in the

\begin{abstract}
Logical query plan rewriting transforms a relational database query into an equivalent but more efficient form and is crucial to the performance of database-backed applications. In existing systems, rewrite rules are typically implemented manually, tightly coupled to specific execution engines, and often lack formal correctness guarantees. Consequently, developing a new engine requires reimplementing both legacy and new rules, incurring significant engineering cost, limiting portability, and every new implementation is an opportunity for introducing new bugs.

We introduce \rulescript, an engine-agnostic domain-specific language (DSL) for developing query rewrite rules. \rulescript separates rule definition from execution infrastructure via a relational algebra-inspired core language, and an explicit decomposition of rules into matching and transformation phases. Developers express rewrites by pattern-matching query plans using \rulescript's core operators and constructing semantically equivalent transformed plans, with all rewrites automatically verified formally to ensure correctness. \rulescript is also extensible: users can define custom operators in terms of the core language to capture engine-specific semantics. To integrate with an existing system, developers need only implement a lightweight adapter that maps \rulescript's core and custom operators to the operators implemented in the target engine.

We evaluate \rulescript by reimplementing 33 rewrite rules from Apache Calcite and extending the language with several custom operators. To demonstrate portability, we automatically deploy these rules to CockroachDB and Apache Data Fusion, two engines with substantially different backends. Our results show that \rulescript enables ``write once, deploy everywhere'' paradigm for query plan rewriting, with minimal effort required to deploy previously written rules on a new data engine.
\end{abstract}

\maketitle

\pagestyle{\vldbpagestyle}
\begingroup\small\noindent\raggedright\textbf{PVLDB Reference Format:}\\
\vldbauthors. \vldbtitle. PVLDB, \vldbvolume(\vldbissue): \vldbpages, \vldbyear.\\
\href{https://doi.org/\vldbdoi}{doi:\vldbdoi}
\endgroup
\begingroup
\renewcommand\thefootnote{}\footnote{\noindent
This work is licensed under the Creative Commons BY-NC-ND 4.0 International License. Visit \url{https://creativecommons.org/licenses/by-nc-nd/4.0/} to view a copy of this license. For any use beyond those covered by this license, obtain permission by emailing \href{mailto:info@vldb.org}{info@vldb.org}. Copyright is held by the owner/author(s). Publication rights licensed to the VLDB Endowment. \\
\raggedright Proceedings of the VLDB Endowment, Vol. \vldbvolume, No. \vldbissue\ %
ISSN 2150-8097. \\
\href{https://doi.org/\vldbdoi}{doi:\vldbdoi} \\
}\addtocounter{footnote}{-1}\endgroup

\ifdefempty{\vldbavailabilityurl}{}{
\vspace{.3cm}
\begingroup\small\noindent\raggedright\textbf{PVLDB Artifact Availability:}\\
The source code, data, and/or other artifacts have been made available at \url{\vldbavailabilityurl}.
\endgroup
}

\section{Introduction}

Query processing is the task of executing data operations that conform to the semantics specified by a user’s query. Central to this process is query optimization, whose objective is to identify the most efficient execution plan among all plans that are semantically equivalent to the input query. Given the vast search space of possible plans, modern optimizers rely heavily on query transformations to improve performance while preserving correctness.

Since the seminal \systemR project~\cite{astrahan1976system, selinger1979access} and the development of extensible optimizer frameworks such as Volcano and Cascades~\cite{graefe1993volcano, graefe1995cascades}, logical plan rewriting has been a foundational technique in relational query optimization. Logical rewriting is driven by \emph{query rewrite rules}, each of which matches a specific structural pattern in a query plan and transforms it into a semantically equivalent but potentially more efficient plan. For example, a rule may detect a selection operator applied after a join and rewrite the plan by pushing the selection below the join, thereby reducing the size of intermediate results and lowering execution cost. Over time, as optimization techniques have evolved, the number and sophistication of rewrite rules in production systems have also grown substantially.

This growth, however, has introduced significant engineering challenges. Modern engines often maintain hundreds of rewrite rules. For instance, Apache \calcite~\cite{begoli2018apache} contains more than one hundred logical rewrite rules, each implemented in imperative \java code and tightly integrated with the optimizer’s internal representation. Similarly, the commercial \cockroach database~\cite{taft2020cockroachdb} includes more than two hundred rewrite rules expressed in its own domain-specific language (DSL). As the rule base expands, so does the cost of development, maintenance, and portability. When building a new engine or refactoring an existing one, developers must frequently reimplement existing rules against a new backend representation, a process that is both time-consuming and error-prone.

A more fundamental issue concerns correctness. Rewrite rules are intended to be semantics-preserving: for every possible database instance, the rewritten plan should produce the same result as the original plan. In practice, however, rewrite rules in major database systems are rarely accompanied by formal proofs of correctness, and subtle bugs have repeatedly arisen from incorrect transformations~\cite{mysql2024bug, sqlite2025bug, postgres2015bug, rigger2020detect}. Automated query equivalence solvers, such as \cosette, \hottsql, QED, and SPES~\cite{chu2017cosette, chu2017hottsql, wang2024qed, zhou2022spes}, offer a principled way to verify semantic equivalence. One approach is to invoke such solvers on every optimized query, comparing the original and rewritten plans. Unfortunately, this strategy incurs significant runtime overhead, as equivalence checking can be computationally expensive and unsuitable for use within the tight latency constraints of query optimization.

A more scalable alternative is to verify {\em the rewrite rules themselves rather than each rewritten query instance}, and the \hottsql language~\cite{chu2017hottsql} represents an important step in this direction. It provides a DSL based on \sql for expressing query semantics and integrates with the \cosette solver~\cite{chu2017cosette} to verify equivalence between plans written in the language. While effective in certain settings, \hottsql has notable limitations: incomplete support for \sql features such as \op{Null} semantics and integrity constraints restricts its coverage, and it is not integrated with production data engines. As a result, developers must manually translate the verified rules into engine-specific implementations, reintroducing duplication and potential inconsistencies.

On the other hand, systems such as \cockroach provide their own DSLs for specifying rewrite rules and can automatically generate corresponding implementations~\cite{cockroach2026optgen}. However, these DSLs are tightly coupled to a single engine and lack formal verification guarantees. Rules written in such languages cannot be easily reused across systems, nor can their correctness be formally assured.

In this paper, we introduce \rulescript, an extensible, verifiable, and engine-agnostic rule language for logical query plan rewriting. The core syntax of \rulescript is based on relational algebra operators such as projection, selection, and join. Unlike concrete query plans derived from executable \sql statements, \rulescript plans may contain \emph{uninterpreted symbols}, including uninterpreted relations, functions, and types. These symbols act as typed placeholders that can be instantiated with arbitrary concrete plans or expressions of the appropriate form. For example, an uninterpreted predicate may represent any Boolean expression over a given schema, while an uninterpreted relation may denote any plan that produces tuples of a specified structure. This abstraction enables a single \rulescript rule to succinctly describe an entire family of concrete rewrites.

All rewrite rules expressed in \rulescript are formally verified. During verification, the system reasons universally over all valid instantiations of uninterpreted symbols. A rule is accepted only if semantic equivalence holds for every possible assignment, ensuring that no concrete instantiation can violate correctness. In this way, \rulescript provides strong, rule-level correctness guarantees without requiring equivalence checks during query optimization at runtime.

To support engine independence, each \rulescript rule consists of two query plans: a \emph{match pattern} and a \emph{transform pattern}. Applying a rule to a concrete plan requires identifying an instantiation of uninterpreted symbols under which the match pattern corresponds to a subplan of the concrete query. If such an instantiation exists, the plan is rewritten by replacing the matched subplan with the transform pattern instantiated under the same assignment. This separation of specification and execution enables a clear and portable rule definition.

\rulescript is also designed to be extensible. In addition to its core operators, users may define {\em custom operators} in terms of the core language to capture engine-specific constructs, such as specialized physical operators or proprietary scalar expressions. These custom operators can then be used within rewrite rules while retaining the ability to verify them through their core-language definitions.

Integrating the verified rules into a concrete data engine is achieved through a lightweight {\em adapter}. The adapter maps \rulescript's core and custom operators to the target engine’s internal representations and APIs. During deployment, the adapter traverses the match and transform patterns and generates executable implementations of verified rules. Once generated, these implementations can be incorporated into the optimizer without incurring any verification overhead at runtime.

In summary, this paper makes the following contributions:
\begin{itemize}
    \item We design \rulescript, an extensible and formally verified DSL for expressing logical query plan rewrite rules. By leveraging uninterpreted symbols, \rulescript enables concise specification of broad classes of semantics-preserving transformations.
    \item We introduce an engine-agnostic architecture based on adapters that map \rulescript operators to concrete optimizer implementations, enabling portable rule deployment across heterogeneous systems.
    \item We implement a prototype of \rulescript and use it to implement 33 rewrite rules from the Apache \calcite engine. All implemented rules are formally verified, and we demonstrate portability by deploying them across multiple real-world backend engines, using 3$\times$ fewer lines of code when compared to the same implementation in \calcite.
\end{itemize}

\section{Overview} \label{sec:motivation}

\begin{figure}[t]
\usetikzlibrary{positioning,fit,arrows.meta,calc}

\begin{tikzpicture}[
  font=\normalsize,
  box/.style={draw, align=center, inner sep=4pt},
  group/.style={draw, dashed, inner sep=4pt},
  arr/.style={-Latex, line width=0.8pt}
]

\node[box] (rule) {Rule specification\\
  $q_{\text{from}} \Rightarrow q_{\text{to}}$ (\autoref{sec:core}, \autoref{sec:extension})
};

\node[box, above left=0mm and 4mm of rule] (eqprob) {Equivalence problem};
\node[box, below=4mm of eqprob] (eqcheck) {Equivalence checker\\(QED~\cite{wang2024qed}, etc.)};

\draw[arr] (eqprob) -- (eqcheck);

\node[group, fit=(eqprob)(eqcheck)] (gleft) {};
\node[anchor=south west] at (gleft.north west){Verification (\autoref{sec:verification})};

\node[box, below=10mm of eqcheck] (match) {Match $q_{\text{from}}$};
\node[box, below=6mm of match] (inst) {Instantiate $q_{\text{to}}$\\with $\sigma$};

\draw[arr] (match) -- node[right, xshift=1mm] {symbols instantiation $\sigma$} (inst);

\node[right=6mm of match] (input) {Concrete query plan};
\node[right=6mm of inst] (output) {Transformed query plan};

\draw[arr] (input) -- (match);
\draw[arr] (inst) -- (output);

\node[group, fit=(match)(inst)(input)(output)] (gright) {};
\node[anchor=south west] at (gright.north west){Execution (\autoref{sec:execution})};

\draw[arr] (rule) |- node[above] {semantic reduction} (eqprob);
\draw[arr] ($(rule.south)+(-14mm,0mm)$) -- node[right, align=left, xshift=1mm] {Adapters targeting\\query engines} (gright);
\end{tikzpicture}

\caption{The full lifecycle of a rewrite rule in \rulescript: Rules specified using the \rulescript DSL (\autoref{sec:core}, \autoref{sec:extension}) can be statically verified for correctness (\autoref{sec:verification}), and executed in query engines through \rulescript adapters (\autoref{sec:execution}).}
\label{fig:algorithm-overview}
\end{figure}
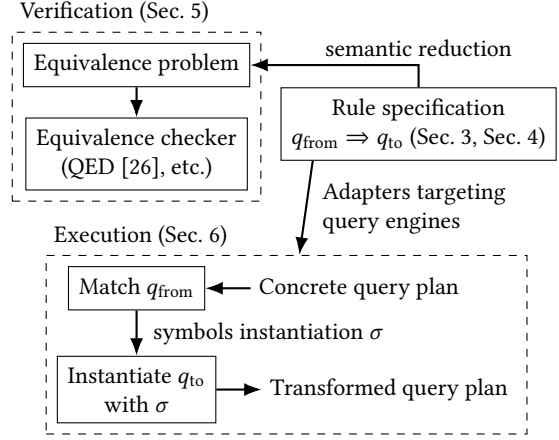

We present a motivating example that illustrates the challenges of expressing and maintaining a general query rewrite rule in a data processing engine, and how \rulescript simplifies that task.

Consider an order management system where we query the total amount spent by each customer who has ever left a 1-star review. This requires aggregation over the \texttt{Order} table and filtering by the existence of bad reviews in the \texttt{Rev} table. In many query engines, such an existence filter is supported directly by a better-optimized \op{SemiJoin} operator, resulting in the following query plan:
\begin{equation} \label{eq:example-plan-before}
\begin{aligned}
&\op{SemiJoin}(\mathit{on}{:}\; \mathit{Order.cust} = \mathit{Rev.author} \land \mathit{Rev.rating} \leq 1, \\
&\quad \op{Aggregate}(\mathit{key}{:}\; \mathit{Order.cust},\; \mathit{val}{:}\; \op{Sum}(\mathit{Order.amt}), \\
&\qquad \op{Scan}(\mathit{Order})), \\
&\quad \op{Scan}(\mathit{Rev}))
\end{aligned}
\end{equation}
Moreover, modern query optimizers~\cite{spark2022rule, doris2022rule} are able to inspect the join condition's structure and push the \op{SemiJoin} through the \op{Aggregate}. This results in an optimized plan with smaller data cardinality during execution:
\begin{equation} \label{eq:example-plan-after}
\begin{aligned}
&\op{Aggregate}(\mathit{key}{:}\; \mathit{Order.cust},\; \mathit{val}{:}\; \op{Sum}(\mathit{Order.amt}), \\
&\quad \op{SemiJoin}(\mathit{on}{:}\; \mathit{Order.cust} = \mathit{Rev.author} \land \mathit{Rev.rating} \leq 1, \\
&\qquad \op{Scan}(\mathit{Order}),\; \op{Scan}(\mathit{Rev})))
\end{aligned}
\end{equation}

However, implementing and maintaining the correctness of such transformations is tedious and challenging:

\paragraph{Challenge 1: Pattern generality}
For maximal generality, most rewrites can be expressed as transforming \emph{patterns} of query plans without hardwiring them to only match on specific schemas or how the input tables are computed. But simply creating a rewrite rule for every possible type of query nodes is unsound: in our example, the operators commute only when the \op{SemiJoin}'s condition does not depend on columns in the left input that are not grouped by the \op{Aggregate}.

\paragraph{Challenge 2: Custom operators}
Query engines usually implement new query operators beyond the standard relational algebra primitives, so that they can optimize better for their execution backends. \op{SemiJoin}s, for example, are known to enable better optimization for parallel and distributed query processing~\cite{valduriez1984semijoin,chen1993combine}. They are syntactically similar to inner joins yet semantically different, so many new potential rewrites specific to \op{SemiJoin}s have to be reasoned through and implemented.

Indeed, the rewrite shown is absent from all 3 database engines that we integrate \rulescript with in \autoref{sec:eval}, and we are only aware of Apache Spark and Doris implementing similar optimizations. Specifically, the Doris implementation~\cite{doris2022rule} overly restricts the form of group-by keys, rendering it less general than the \rulescript rule in \autoref{eq:semijoin-agg-rule}. The Spark implementation~\cite{spark2022rule} has similar generality but with higher code complexity: requiring 66 lines of Scala code to express the same rule as \autoref{eq:semijoin-agg-rule}.

\medskip

To address Challenge~1, we show how \rulescript captures this optimization rule in its most general form while maintaining soundness using only a few lines of code. In the pattern language to be presented in \autoref{sec:core}, our example rule can be written as:
\begin{equation} \label{eq:semijoin-agg-rule}
\begin{aligned}
q_\text{from} &= \op{SemiJoin}(\lambda (k, v)\, y.\; P(k, y), \\[-2pt]
&\qquad\qquad \op{Aggregate}(\lambda x.\; G(x),\; \lambda x.\; A(x),\; L),\; R) \\[4pt]
q_\text{to} &= \op{Aggregate}(\lambda x.\; G(x),\; \lambda x.\; A(x), \\[-2pt]
&\qquad\qquad \op{SemiJoin}(\lambda x\, y.\; P(G(x), y),\; L,\; R))
\end{aligned}
\end{equation}
where \(q_\text{from}\) and \(q_\text{to}\) are the \emph{patterns} of query plans before and after the transformation. 

The left and right inputs of \op{SemiJoin} can be any sub-query of any schema, so we introduce two uninterpreted sub-query \emph{plan symbols} \(L : \op{Bag}\langle X \rangle\) and \(R : \op{Bag}\langle Y \rangle\) with generic row types \(X\) and \(Y\). The \op{Aggregate} operator partitions input rows into groups by a key function and aggregates each group into a single value row. We allow the key and aggregation to be unconstrained by introducing the uninterpreted \emph{function symbol} \(G : X \to K\) into some generic key type \(K\), and the aggregation function symbol \(A : \op{Bag}\langle X \rangle \to V\) into some generic value type \(V\).

The \op{SemiJoin}'s condition is more subtle. In \(q_\text{from}\), the join condition ranges over the left inputs of type \((K, V)\) from the \op{Aggregate} and the right inputs of type \(Y\). For the rule to be correct, its join condition must not depend on the non-key columns of the \op{Aggregate}. This constraint is encoded by introducing the \emph{predicate symbol} \(P : (K, Y) \to \op{Bool}\) and constraining the condition to the form \(\lambda (k, v)\, y.\; P(k, y)\), which drops the non-key column \(v\). In \(q_\text{to}\), the \op{SemiJoin} operates over the un-aggregated left inputs of type \(X\), yet the join condition should still be over the aggregate key: we encode this as \(\lambda x\, y.\; P(G(x), y)\), applying \(G\) to transform the left input before filtering with \(P\). Thus, our \rulescript rule ensures maximal generality by only including minimal constraints to ensure semantic equivalence between the \(q_\text{from}\) and  \(q_\text{to}\), while not restricting the schemas of the input relations and allowing them to be derived using any operator.

To address Challenge~2, \rulescript allows users to easily declare new operators by describing the signature of custom operators and explicitly modeling them using \rulescript's core relational semantics to be described in \autoref{sec:extension}. For example, we declare that the \op{SemiJoin} operator takes a condition, a left sub-query, and a right sub-query. Its relational semantics is given by translating \op{SemiJoin} into core relational operators:
\begin{equation} \label{eq:semijoin-def}
\begin{aligned}
&\op{SemiJoin}(\lambda x\, y.\; P(x, y),\; L,\; R) \\
&\quad := \op{Filter}(\lambda x.\; \exists(\op{Filter}(\lambda y.\; P(x, y),\; R)),\; L)
\end{aligned}
\end{equation}
Here the join condition $P$ takes in two parameters $x$ and $y$, referring to rows from the left and right sub-queries $L$ and $R$.
This allows \op{SemiJoin} in the rewrite rule to be matched and transformed as one unit, while being \emph{semantically transparent} to our verification process without increasing the trust surface.

In \autoref{sec:verification}, we present the verification of rewrite rules in \rulescript that accounts for of the full generality of query patterns and custom operators. Concretely, \rulescript translates the rule in \autoref{eq:semijoin-agg-rule} into a query equivalence problem that a) keep all uninterpreted symbols uninterpreted, and b) substitute all custom operators with their core-relational equivalent.
Then \rulescript presents to the \qedp prover~\cite{wang2024qed} the equivalence problem $q_A \overset{?}{=} q_B$:
\begin{equation}
\begin{aligned}
q_A &= \op{Filter}(\lambda (k, v).\; \exists(\op{Filter}(\lambda y.\; P(k, y),\; R)),  \\[-2pt]
&\qquad\qquad \op{Aggregate}(\lambda x.\; G(x),\; \lambda x.\; A(x),\; L)) \\[4pt]
q_B &= \op{Aggregate}(\lambda x.\; G(x),\; \lambda x.\; A(x), \\[-2pt]
&\qquad\qquad \op{Filter}(\lambda x.\; \exists(\op{Filter}(\lambda y.\; P(G(x), y),\; R)),\; L)
\end{aligned}
\end{equation}

After verification, \rulescript further provides adapters that users can implement to target different query optimization engines ensuring \emph{sound} execution of the declared and verified rewrite rules.
To be described in \autoref{sec:execution}, the execution of a rule is a two-step pipeline:
1) Given a concrete query plan $q$, we attempt to \emph{match} it with the $q_\text{from}$ pattern of the rewrite rule and see if the rule is applicable.
If so, the match will also synthesize a instantiation of all uninterpreted symbols in $q_\text{from}$ with concrete expressions in $q$.
2) The instantiation is then applied to the $q_\text{to}$ pattern and yields a concrete transformed query plan.

For example, matching the concrete query plan \autoref{eq:example-plan-before} with $q_\text{from}$ results in the following instantiation of symbols:
\begin{equation*}
\begin{aligned}
P &\mapsto \lambda (k,y).\; k.\mathit{cust} = y.\mathit{author} \;\land\; y.\mathit{rating} \leq 1 \\
G &\mapsto \lambda x.\; x.\mathit{cust} \qquad
A \mapsto \lambda x.\; \op{Sum}(x.\mathit{amt}) \\
L &\mapsto \op{Scan}(\mathit{Order}) \qquad
R \mapsto \op{Scan}(\mathit{Rev})
\end{aligned}
\end{equation*}
Applying this instantiation to \(q_\text{to}\) then gives the transformed query plan \autoref{eq:example-plan-after}. Note that there is no need to verify the semantic equivalence on the concrete query plans, since \rulescript already checks the rule is correct for \emph{all} instantiations of the uninterpreted symbols in the previous rule-level verification.

To be discussed in \autoref{sec:eval}, to test the real-world effectiveness and portability of \rulescript,
we translated 33 rewrite rules from the Apache Calcite engine into the \rulescript specification language. We also implemented \rulescript \emph{adapters} for 3 different real-world database engines, and demonstrated the same 33 \rulescript rules can be registered and executed in all of them, using up to 3$\times$ fewer lines of code compared to the original Calcite implementation.

\section{\rulescript's Core Language} \label{sec:core}
We now present the core language for \rulescript. Optimization rules match and transform plans by their operator skeleton, so \rulescript's syntax mirrors the structure of relational plans found in common database engines. To make rules schema-agnostic, the language leaves types, expressions, and subplans within that skeleton as uninterpreted symbols. Matching a rule reduces to finding an instantiation of these symbols; applying the rule evaluates the transform pattern under the same instantiation. We present the syntax of \rulescript patterns in \autoref{sec:syntax} and formalize it in \autoref{sec:interpret}.

\subsection{Syntax} \label{sec:syntax}
To distinguish a \rulescript expression from a concrete query plan, we refer to the former as a \emph{pattern}. \autoref{fig:syntax} gives the full grammar of \rulescript. A rewrite rule is a pair of patterns \((q_\text{from}, q_\text{to})\): a \emph{match pattern} describing which query plans the rule applies to, and a \emph{transform pattern} describing the result of the rewrite once matched. Optionally, a rule may include a first-order constraint on its uninterpreted symbols (e.g., requiring a function to be injective); we omit this when no constraint is needed.

\begin{figure}[t]
\centering
\(\displaystyle
\begin{array}{rcl}
q &::=& Q : \op{Bag}\langle\sigma\rangle
  \mid \op{Empty}(\sigma)
  \mid \op{Filter}(\lambda x.\; p,\; q) \\
  &\mid& \op{Project}(\lambda x.\; e,\; q)
  \mid \op{Join}(\lambda x\, y.\; p,\; q,\; q) \\
  &\mid& \op{Union}(q,\; q)
  \mid \op{Distinct}(q) \\
  &\mid& \op{Aggregate}(\lambda x.\; e,\; \lambda x.\; \alpha,\; q) \\[6pt]
p &::=& e \mid \top \mid \bot \mid \lnot p \mid p \land p \mid p \lor p \\
  &\mid& e = e \mid e \neq e \mid e\;\op{is\;null} \mid e\;\op{is\;not\;null} \\
  &\mid& \exists(q) \\[6pt]
e &::=& x \mid c \mid f(e, \ldots) \\[6pt]
\alpha &::=& g(e, \ldots) \\[4pt]
\end{array}
\)

\smallskip
\begin{tabular}{rl}
\text{where} & \(f : \sigma \to \sigma\) \quad\text{(function symbol),} \\
& \(g : \op{Bag}\langle\sigma\rangle \to \sigma\) \quad\text{(aggregate function symbol),} \\
& \(\sigma\) \quad\text{(type symbol)}
\end{tabular}
\caption{Syntax of \rulescript patterns. Here \(q\) ranges over patterns, \(p\) over predicates, \(e\) over scalar expressions, and \(\alpha\) over aggregate expressions. Lambda binders \(\lambda x\) name the input tuple, making column references explicit.}
\label{fig:syntax}
\end{figure}

Uninterpreted symbols serve dual roles. During \emph{verification}, \rulescript checks that the match and transform patterns produce equivalent plans for \emph{all} possible instantiations, establishing correctness regardless of the concrete plans and expressions a rule is applied to. During \emph{execution}, the system only needs to find \emph{one} instantiation that makes the match pattern agree with a given concrete plan; it then evaluates the transform pattern under the same instantiation. There are three kinds of uninterpreted symbols.

\paragraph{Uninterpreted types}
A type symbol such as \(X\) or \(K\) stands for any concrete value type. We require every type to support equality comparison and to contain a distinguished \op{Null} value. Type symbols with the same name are considered identical across a rule.

Crucially, a type symbol represents the \emph{entire} tuple type of a plan's output, not an individual column. For example, instantiating \(X\) to \(\op{Int} \times \op{String}\) lets a single symbol represent a two-column schema. Column-level structure is not part of the abstract type; instead, uninterpreted functions access individual fields. For instance, given \(L : \op{Bag}\langle X \rangle\) and \(G : X \to K\), the function \(G\) can be instantiated to a projection that extracts a single column, to the identity function that returns the entire tuple, or to multiple arithmetic expressions over different columns. The abstract type \(X\) does not prescribe how many columns exist or how they are named.

\paragraph{Uninterpreted functions}
A function symbol such as \(P : (K, Y) \to \op{Bool}\) or \(G : X \to K\) stands for any function with the declared signature. It can be instantiated by any concrete expression with matching types: for example, column references, arithmetic, or comparisons. An \emph{aggregate} function symbol such as \(A : \op{Bag}\langle X \rangle \to V\) is distinguished by its input type: it maps a bag of tuples to a single value, and can be instantiated by aggregate operators like \(\op{Sum}\) or \(\op{Count}\). Function symbols with the same name are considered identical across a rule.

\paragraph{Uninterpreted plans}
A plan symbol such as \(L : \op{Bag}\langle X \rangle\) stands for any concrete query plan whose output tuples have a type matching \(X\) (under some instantiation of the type symbols). Plan symbols are the leaf nodes of a pattern. The symbol \(\op{Empty}(\sigma)\) is a special case: it matches only plans producing no tuples, but still carries a type to keep the schema well-defined.

\subsection{Interpretation} \label{sec:interpret}
A pattern represents a \emph{family} of concrete query plans: all those obtainable by choosing concrete types, functions, and input plans for its uninterpreted symbols. We call such a choice an \emph{instantiation}, and any concrete plan that results from some instantiation a \emph{valid interpretation} of the pattern.

Applying a rewrite rule \((q_\text{from}, q_\text{to})\) to a concrete query plan proceeds in two steps:
\begin{enumerate}
    \item \textbf{Match.} Find an instantiation of the uninterpreted symbols in the rewrite rule such that \(q_\text{from}\) evaluates to the given concrete plan.
    \item \textbf{Transform.} Evaluate \(q_\text{to}\) under the \emph{same} instantiation to produce the rewritten plan.
\end{enumerate}
Verification ensures that \emph{every} instantiation yields an equivalent pair of plans, hence every rule only needs to be verified once. Execution only requires finding \emph{some} instantiation for which the match pattern agrees with the concrete plan; a valid instantiation that the execution engine does not discover simply means the rule does not fire, so that correctness is unaffected.

We now describe how each pattern variant is interpreted. The lambda binder \(\lambda x\) names the tuple flowing through the operator, making column references and dependencies explicit.

\subsubsection{Plan symbols}
A symbol \(L : \op{Bag}\langle X \rangle\) matches any concrete query plan whose output type equals the concrete type assigned to \(X\). As noted above, because \(X\) is uninterpreted, a single symbol can represent an arbitrarily concrete schema consisting of many columns. In a match pattern, encountering a plan symbol records the concrete subplan and binds the symbol to it. In a transform pattern, the same symbol refers back to the captured subplan. \(\op{Empty}(\sigma)\) behaves identically except that it only matches plans that produce no tuples.

\subsubsection{Recursive operators}
The remaining variants are defined recursively: each has one or more child patterns and possibly additional arguments.

\paragraph{\op{Filter} and \op{Join}}
A pattern \(\op{Filter}(\lambda x.\; P(x),\; q)\) represents any concrete \(\op{Filter}\) whose input is a valid interpretation of \(q\) and whose predicate is a valid interpretation of \(\lambda x.\; P(x)\). Here \(P\) is an uninterpreted predicate that can be instantiated with any Boolean expression over the input columns.

A valid interpretation needs not be syntactically similar to the pattern. For example, \(\op{Filter}(\lambda x.\; P_0(x) \land P_1(x),\; q)\) can be satisfied by a concrete filter whose predicate is \(\mathtt{a > 5\ AND\ b < 10}\): i.e., we assign \(P_0(x) \mapsto x.a > 5\) and \(P_1(x) \mapsto x.b < 10\). While verification reasons over all such assignments; execution only needs to find one assignment by partitioning the concrete predicate's conjuncts among the pattern's predicate symbols.

\(\op{Join}(\lambda x\, y.\; P(x, y),\; q_0,\; q_1)\) extends \(\op{Filter}\) with a second input pattern. Its predicate ranges over tuples from both inputs. In the core language, \(\op{Join}\) denotes an inner join; other join types are introduced by defining them as custom operators to be discussed in~\autoref{sec:extension}.

\paragraph{\op{Project} and \op{Aggregate}}
A pattern \(\op{Project}(\lambda x.\; F(x),\; q)\) represents any concrete \(\op{Project}\) whose input matches \(q\) and whose projection function is a valid interpretation of \(F\). Here \(F : X \to Y\) is an uninterpreted function mapping input tuples to output tuples.

\(\op{Aggregate}(\lambda x.\; G(x),\; \lambda x.\; A(x),\; q)\) takes two function arguments: a grouping key and an aggregation. The key function \(G : X \to K\) extracts the grouping columns from each input tuple. The aggregation \(A : \op{Bag}\langle X \rangle \to V\) is distinguished by its type: it maps the \emph{bag} of tuples sharing the same key to a single output value. This type distinction is necessary because \(A\) can be instantiated by operators like \(\op{Sum}\) or \(\op{Count}\), which consume an entire group rather than a single tuple. The output type of an \(\op{Aggregate}\) is \(K \times V\): one component from the key and one from the aggregation.

Referring back to the running example from \autoref{sec:motivation}, the pattern \(\op{Aggregate}(\lambda x.\; G(x),\; \lambda x.\; A(x),\; L)\) matches a concrete aggregation that groups orders by customer and sums their amounts, with \(G\) instantiated to the customer column extraction and \(A\) instantiated to \(\op{Sum}\) over the amount column.

\paragraph{\op{Union} and \op{Distinct}}
These are the simplest recursive operators. \(\op{Union}\) has two child patterns and no additional arguments; \(\op{Distinct}\) has one. They match concrete plans of the same kind and recursively verify their inputs.

\subsubsection{Custom operators} \label{sec:core:alias}
The core language covers standard relational operators, but real-world optimizers often define additional operators, such as semi-joins or anti-joins, as first-class constructs. \rulescript accommodates this through an extension mechanism as mentioned earlier: users define custom operators whose semantics are expressed in terms of the \rulescript's core operators. For verification, the custom operator is expanded to its core definition; for execution, users provide handlers that match and transform the operator directly. We describe this mechanism in \autoref{sec:extension}.

\section{Extending \rulescript} \label{sec:extension}

The core language presented in \autoref{sec:core} supports standard relational operators. However, real-world query optimizers often define additional operators tailored to their execution engines. For example, DataFusion, CockroachDB, and Calcite all support a dedicated \op{SemiJoin} operator that is implemented more efficiently than expressing the same semantics through standard operators. To write rewrite rules involving such operators, \rulescript provides an extension mechanism that lets users define custom operators and use them in rules alongside the core operators.

\rulescript's extension design rests on two principles that address verification and execution separately.

First, users define a custom operator by providing its \emph{semantics} expressed entirely using \rulescript's core operators (or other defined custom operators). This is both necessary and sufficient for verification: \rulescript expands each custom operator into its core operator-only definition, yielding patterns that solvers can reason about directly (\rulescript currently uses the \qedp solver, to be discussed in~\autoref{sec:verification}). As long as the semantics faithfully captures the operator's meaning, the correctness proof carries over to any rule that uses it.

Second, during execution the custom operator remains \emph{opaque}. The system matches and transforms it as a single node, without expanding it to the (potentially less efficient) core operator-only form. This is necessary because a backend's native representation of the custom operator may differ substantially from the semantic definition due to implementation-level optimizations. To execute rules involving custom operators, users provide a \emph{handler}: a short, backend-specific function that recognizes a custom operator node in a concrete plan and extracts its parameters, including child plans, predicates, and expressions. The handler encapsulates only the structural knowledge specific to the custom operator: identifying the node type and locating each parameter within the backend's plan representation. Once the parameters are extracted, \rulescript's pipeline matches and instantiates them using the same mechanism as for core operators (\autoref{sec:interpret}), since the parameter types, such as plans, predicates, scalar expressions, aggregates, are the same regardless of which operator contains them. 

\subsection{Definition} \label{sec:extension:def}

To define a custom operator, users provide four components: the operator's \emph{name}, its \emph{typed parameters} (child plans and any additional arguments such as predicates), its \emph{output type}, and its \emph{semantics} expressed as a core operator-only pattern. \autoref{fig:extension-syntax} provides the syntax for defining custom operators.

\begin{figure}[t]
\centering
\(\displaystyle
\begin{array}{rcl}
d &::=& \op{op}(\overline{a}) : \op{Bag}\langle\sigma\rangle := q \\[4pt]
a &::=& Q : \op{Bag}\langle\sigma\rangle
  \mid \lambda \overline{x}.\; p
  \mid \lambda \overline{x}.\; e
  \mid \lambda \overline{x}.\; \alpha \\
\end{array}
\)
\caption{Syntax for custom operator definitions. Here \(d\) is a definition, \(a\) ranges over parameters (child plans, predicates, scalar expressions, or aggregate expressions), \(\overline{a}\) is an ordered list of parameters, and \(q\) is a core-operator pattern giving the semantics. The notation \(\overline{x}\) denotes a list of bound variables.}
\label{fig:extension-syntax}
\end{figure}

We illustrate with the \op{SemiJoin} operator from \autoref{sec:motivation}. Its definition (\autoref{eq:semijoin-def}) has three parameters: two input plans \(L : \op{Bag}\langle X \rangle\) and \(R : \op{Bag}\langle Y \rangle\), and a join predicate \(P : (X, Y) \to \op{Bool}\). The output type is \(\op{Bag}\langle X \rangle\), reflecting that semi-joins preserve only the left input's schema.

\subsection{Usage}

Once defined, a custom operator can be used in rewrite rules exactly like a core operator. The \op{SemiJoin}-Aggregate transpose from \autoref{sec:motivation} (\autoref{eq:semijoin-agg-rule}) states that when the semi-join predicate depends only on the grouping key (not the aggregated value), the aggregation can be pushed below the semi-join. This is beneficial because the semi-join then operates on the unaggregated input, potentially filtering rows before the more expensive aggregation.

\subsection{Verification and Execution}

\paragraph{Verification}
Once defined, \rulescript will attempt to verify each rule before deploying it during runtime. To verify our semi-join rule, \rulescript expands each occurrence of the \op{SemiJoin} operator using its semantics (\autoref{eq:semijoin-def}), yielding match and transform patterns composed entirely of core operators. \rulescript then uses a SQL solver (currently \qedp~\cite{wang2024qed}) to check the equivalence for all instantiations of the uninterpreted symbols, to be described in \autoref{sec:verification}. The correctness guarantee transfers from the expanded core operator-only patterns to the original rule with custom operators, because the semantics definition establishes their equivalence by construction.

\paragraph{Execution}
During execution, the custom operator is not expanded. A handler for a custom operator follows the same pattern as a core operator during matching and transformation. During matching, it recognizes the concrete node, then delegates each parameter to the matching logic for that parameter's type: child plans are matched as plan symbols, predicates as uninterpreted functions, and so on. Each step populates the context with bindings and type instantiations in the same way as the corresponding core operator would. During transformation, the handler retrieves bindings from the context and constructs the output node, again rewriting expressions using the same type instantiation logic as core operators.

For example, the \op{SemiJoin} match handler checks that the concrete node is a semi-join, then invokes the plan matching logic for \(L\) and \(R\) and the predicate matching logic for \(P\), with the same routines used by \op{Filter} and \op{Join}. The transform handler constructs a new semi-join node from the instantiated subplans and predicate. Because the matching and instantiation logic is reused from core operators, adding a new custom operator requires only a small amount of backend-specific code that maps the operator's parameters to their types.

\rulescript's decoupling of verification from execution of custom operators allows it to work directly with a data engine's native implementation of custom operators. For instance, \op{SemiJoin} may be implemented quite differently from the \(\exists\)-based definition in a data engine, yet the verification guarantee of our rewrite rule still holds.

\section{Verification} \label{sec:verification}
Verifying a rewrite rule requires showing equivalence not for one concrete plan, but for \emph{all} instantiations of the uninterpreted symbols. This rules out testing-based approaches and requires a solver that can reason universally over uninterpreted types and functions. \qedp~\cite{wang2024qed} is currently the only automated query equivalence solver with this capability and sufficient coverage of SQL features, making it our choice for verification. To leverage \qedp, we encode \rulescript's match and transform patterns into \qedp's input format. For rules involving custom operators (\autoref{sec:extension}), \rulescript first expands them using their semantics definitions, yielding patterns composed entirely of core operators. We describe \qedp's syntax in \autoref{sec:formalism} and briefly describe how it verifies equivalence in \autoref{sec:qed}.

\subsection{Syntax of Query Plans} \label{sec:formalism}
Here we briefly list the formalism for query plans which are the inputs to the \qedp solver. We list the common syntax and the high-level semantics for them in \autoref{tab:queryplan}. Note that \autoref{tab:queryplan} shows \qedp's input syntax, where predicates and functions are applied positionally (e.g., \(\op{Filter}(P, Q)\)). This corresponds directly to the \(\lambda\)-binder notation used in \rulescript patterns (\autoref{sec:syntax}): the binder \(\lambda x.\; P(x)\) makes variable scope explicit, but the underlying predicate \(P\) is the same object passed to \qedp.

\begin{table}[ht]
    \caption{Syntax for \qedp Query Plans}
    \label{tab:queryplan}
    \begin{tabular}{rp{0.6\linewidth}}
        \toprule
        Notation & High-Level Semantics\\
        \midrule
        \(Q, Q_0, Q_1, \dots\) & Query plans, which output tuples\\
        \(v, v_0, v_1, \dots\) & Individual tuples\\
        \(P, P_0, P_1, \dots\) & Predicates, which map a tuple to a Bool \\
        \(f, f_0, f_1, \dots\) & Functions, which map a tuple to a tuple \\
        \(\alpha, \alpha_0, \alpha_1, \dots\) & Aggregations, which map a collection of tuples to a tuple \\
        \(\exists(Q)\) & True if \(Q\) produces at least one tuple \\
        \(\op{Table}(R: S)\) & A table named \(R\) with schema \(S\)\\
        \(\op{Value}(v_0, \dots, v_n)\) & A table containing values \(v_0, \dots, v_n\) \\
        \(\op{Filter}(P, Q)\) & Filter \(Q\) with \(P\)\\
        \(\op{Project}(f, Q)\) & Map \(Q\) with \(f\)\\
        \(\op{Join}(P, Q_0, Q_1)\) & Inner join \(Q_0\) and \(Q_1\) on \(P\)\\
        \(\op{Union}(Q_0, Q_1)\) & Union \(Q_0\) and \(Q_1\)\\
        \(\op{Intersect}(Q_0, Q_1)\) & Intersect \(Q_0\) and \(Q_1\)\\
        \(\op{Minus}(Q_0, Q_1)\) & Minus/Except \(Q_0\) and \(Q_1\)\\
        \(\op{Aggregate}(\alpha, f, Q)\) & Aggregate \(Q\) with \(\alpha\) grouped by \(f\)\\
        \(\op{Distinct}(Q)\) & Deduplicate \(Q\) \\
        \bottomrule
    \end{tabular}
\end{table}

The output of a query plan is a collection of tuples. We define the type of a tuple as the product type of its element types. For example, the type of \((0, ``s")\) is \(\op{Integer}\times\op{String}\). \qedp adopts bag semantics: query plans produce multisets of tuples, so equivalence accounts for tuple multiplicity but not ordering. This covers all operators in the core language (\autoref{sec:core}); order-dependent operators such as \op{Sort} and \op{Window} are excluded from \rulescript's current scope and discussed in \autoref{sec:eval}.

\subsection{Semi-Ring Semantics for Query Plans}  \label{sec:qed}
The \qedp solver verifies that the pair of \rulescript query plans in the rewrite rules are equivalent for all uninterpreted types and function operators that appear in them. Within the \qedp solver, the query plans are recursively translated into semi-ring expressions. For example, a \(\op{Table}(R: S)\) is translated into a finitely supported function whose input is any tuple that has the same type as the table schema \(S\) and it outputs the multiplicity of the provided tuple in the table, while a \(\op{Filter}(P, Q)\) is the product of an indicator function of the predicate and the semi-ring expression for the input query plan.

After \qedp translates the query plan into semi-ring expressions, it normalizes it with a restricted set of rules into a finite summation of terms, which are unbounded summations of products of indicator functions and finitely supported functions. Given a pair of query plans, \qedp compares their normalized forms term by term, and then tries to find equivalent pairs of terms using the help of \smt solvers. \qedp does not reason about uninterpreted symbols itself, and the support for uninterpreted symbols comes directly from the underlying \smt solvers that \qedp uses. A pair of terms are provably equal if the underlying \smt solver cannot find a counter-example that distinguishes them. If \qedp manages to find matching between the terms from the pair of query plans, then the two query plans are provably equal.

Because \qedp delegates reasoning about uninterpreted symbols to the underlying \smt solver, the encoding preserves the generality of \rulescript's patterns: a successful proof holds for all instantiations, not just the tested ones.

\section{Rule Execution} \label{sec:execution}

To deploy verified \rulescript rules in an existing optimizer, users implement a backend-specific \emph{adapter} that matches rules against concrete query plans and applies the transformation when matched. \rulescript is designed so that this integration is straightforward: rule execution always decomposes into two phases, matching and transformation, that traverse the rule's patterns recursively, accumulating bindings in a shared context. We describe this common structure in \autoref{sec:pipeline}.

We have implemented two execution mechanisms following this structure. An \emph{interpreter} (\autoref{sec:interpreter}) implements the matching and instantiation logic once per operator type, so that adding a new rule requires only its pattern definition with no per-rule code. This is the approach we use for \datafusion. A \emph{code generator} (\autoref{sec:code-generation}) transpiles \rulescript patterns into a backend's native rule format, so that generated rules integrate alongside hand-written ones. This is the approach we use for \calcite and \cockroach, which provide their own rule frameworks (\texttt{RelRule} and \texttt{Optgen} respectively).

\subsection{Execution Pipeline} \label{sec:pipeline}
Both execution mechanisms follow the match-then-transform semantics introduced in \autoref{sec:interpret}. We describe this structure here; \autoref{sec:interpreter} and \autoref{sec:code-generation} show how our interpreter and code generator each realize it. The two phases are connected by a \emph{context} object that stores bindings, and both dispatch to handlers based on the pattern type (e.g., \(\op{Filter}\), \(\op{Join}\), plan symbol).

\paragraph{Context}
The context accumulates two kinds of bindings during the match phase. First, \emph{source bindings} map plan symbols to the concrete subplans they matched. Second, \emph{function bindings} map uninterpreted function and predicate symbols to their concrete instantiations as expressions.

Source bindings additionally carry \emph{type instantiations} that record how each abstract type maps to concrete columns. Because a single uninterpreted type \(X\) may represent a multi-column schema (as described in \autoref{sec:syntax}), the context must track which concrete columns correspond to each abstract type. This mapping can be derived by different strategies. A \emph{positional} strategy matches each uninterpreted type to exactly one concrete column by position. A \emph{dependency-based} strategy examines the expressions that reference the abstract type and collects all concrete columns they depend on. The dependency-based strategy is more general, as it handles cases where one abstract type maps to multiple concrete columns. Our \datafusion interpreter uses the dependency-based strategy; our \calcite code generator uses the positional strategy; and for \cockroach, \texttt{Optgen} handles type resolution internally, so the code generator delegates this to the backend. Both strategies yield valid type instantiations; they differ only in which concrete plans the matcher can successfully match.

To illustrate the dependency-based strategy, consider matching \(\op{Aggregate}(\lambda x.\; G(x),\; \lambda x.\; A(x),\; L)\) from the running example (\autoref{sec:motivation}). When \(L\) binds to \(\op{Scan}(\mathit{Order})\), the type \(X\) maps to the full Order schema. The strategy then examines \(G\)'s concrete expression \(x.\mathit{cust}\) and maps \(K\) to \(\{\mathit{cust}\}\), and examines \(A\)'s concrete expression \(\op{Sum}(x.\mathit{amt})\) and maps \(V\) to \(\{\op{Sum}(\mathit{amt})\}\). Later, when matching the predicate \(P(k, y)\), the matcher uses these mappings to verify that the concrete predicate depends only on columns in \(K\) and \(Y\), not \(V\), enforcing the type constraint \(P : (K, Y) \to \op{Bool}\).

\paragraph{Match phase}
The match phase walks the match pattern alongside the concrete plan top-down. At each node, it checks that the concrete plan has the expected operator type. For example, an \(\op{Aggregate}\) pattern requires a concrete \(\op{Aggregate}\) node. It first recursively matches the child patterns, populating the context with source and type bindings for the subtrees. It then resolves the node's expressions (predicates, projection functions, aggregate functions) against the concrete expressions using the type instantiations established by the matched children.

Predicate matching requires special handling for conjunctions as \rulescript natively supports boolean operators. When the pattern contains \(P_0(x) \land P_1(x)\), the concrete predicate need not have the same syntactic structure. Instead, the matcher flattens both the pattern and the concrete predicate into lists of conjuncts, then partitions the concrete conjuncts among the pattern predicates based on which abstract types their column dependencies fall under. For example, a pattern \(\lambda x\, y.\; P(x) \land Q(y)\) applied to a concrete predicate \(\mathtt{a > 5\ AND\ b = c\ AND\ d < 10}\) would assign \(\mathtt{a > 5}\) and \(\mathtt{d < 10}\) to \(P\) (if \(a\) and \(d\) belong to \(x\)) and \(\mathtt{b = c}\) to \(Q\) (if \(b\) and \(c\) belong to \(y\)). A greedy assignment suffices for a practical implementation, though more sophisticated strategies (e.g., backtracking or constraint-based) could recover additional matches. If no valid partition exists, the rule does not apply.

When the match phase encounters a plan symbol, it records the concrete subplan in the context and establishes type instantiations for its abstract type. If matching succeeds at every node, the phase completes with a fully populated context.

\paragraph{Transform phase}
The transform phase walks the transform pattern bottom-up and constructs the output plan using the bindings captured during matching. When it encounters a plan symbol, it retrieves the corresponding concrete subplan from the context. When it encounters a predicate or function symbol, it retrieves the bound expression and rewrites column references as needed using the type instantiations. The phase recursively builds operators, connecting them according to the transform pattern's structure.

This structure extends naturally to custom operators defined via the extension mechanism (\autoref{sec:extension}). As described in \autoref{sec:extension}, handlers for custom operators follow the same two-phase pattern: during matching, the handler extracts the operator's parameters and delegates each to the matching logic for its type; during transformation, it constructs the output node from the instantiated bindings.

\subsection{Interpreter} \label{sec:interpreter}
The interpreter realizes the execution pipeline by operating directly on a backend's in-memory query plans at runtime. This approach is appropriate when the target backend does not provide its own rule framework with pattern matching or plan construction support. For example, \datafusion exposes only an \texttt{OptimizerRule} trait whose \texttt{rewrite()} method receives a plan node; all matching and construction logic must be supplied by the rule implementation. The interpreter provides this logic once per operator type (e.g., how to match a concrete \texttt{Filter} node, how to construct a new \texttt{Join} node), so that each rule contributes only a pair of patterns \((q_\text{from}, q_\text{to})\) with no per-rule implementation code. In our \datafusion implementation, each rule implements a \texttt{
RewriteRule} trait by providing a \texttt{from()} and \texttt{to()} method that 
return pattern ASTs; the interpreter handles everything else.

We illustrate with the \op{SemiJoin}-Aggregate transpose rule (\autoref{eq:semijoin-agg-rule}) applied to a plan that computes total spending per customer, filtered to customers with a 1-star review as described in~\autoref{sec:motivation}.

\paragraph{Match phase}
Matching proceeds top-down against \(q_\text{from}\):
\begin{enumerate}
    \item The \op{SemiJoin} handler recognizes the concrete node and exposes its predicate, left input, and right input.
    \item The left child \(\op{Aggregate}(\lambda x.\; G(x),\; \lambda x.\; A(x),\; L)\) is matched recursively: first \(L\) binds to \(\op{Scan}(\mathit{Order})\) with \(X\) instantiated to the Order schema, then \(G \mapsto \lambda x.\; x.\mathit{cust}\) and \(A \mapsto \lambda x.\; \op{Sum}(x.\mathit{amt})\).
    \item The right child \(R\) binds to \(\op{Scan}(\mathit{Rev})\).
    \item The predicate \(\lambda (k,v)\, y.\; P(k, y)\) is resolved: the concrete predicate depends on \(k\) and \(y\) but not \(v\), consistent with \(P : (K,Y) \to \op{Bool}\). So \(P \mapsto \lambda (k,y).\; k.\mathit{cust} = y.\mathit{author} \land y.\mathit{rating} \leq 1\).
\end{enumerate}

\paragraph{Transform phase}
The transform phase constructs \(q_\text{to}\) bottom-up. The key step is building the inner semi-join predicate from the pattern \(\lambda x\, y.\; P(G(x), y)\): substituting the binding for \(G\) replaces \(G(x)\) with \(x.\mathit{cust}\), which is then passed as \(P\)'s first argument, yielding \(\lambda x\, y.\; x.\mathit{cust} = y.\mathit{author} \land y.\mathit{rating} \leq 1\). The remaining bindings are substituted directly, producing the optimized plan where the semi-join filters rows \emph{before} aggregation.

\medskip

To integrate with an existing optimizer, each \rulescript rule is wrapped in a thin adapter that implements the backend's native rule interface. In our \datafusion implementation, each rule is wrapped in a \texttt{RuleWrapper} struct that implements \datafusion's \texttt{OptimizerRule} trait. The optimizer schedules these wrapped rules alongside hand-written ones, traversing the plan tree and applying each rule wherever it matches.

\subsection{Code Generation} \label{sec:code-generation}
When the target backend already provides a rule framework that handles plan traversal, rule scheduling, and plan construction, it is more practical to generate code that plugs into this framework than to reimplement its infrastructure. For example, \calcite's \texttt{RelRule} framework provides skeleton-based pattern matching and a \texttt{RelBuilder} API for constructing plans, while \cockroach's \texttt{Optgen} provides a declarative pattern-rewrite language that compiles to Go. In both cases, a code generator transpiles \rulescript patterns into the framework's native format, and the generated rules integrate directly with the backend's existing toolchain alongside hand-written rules.

This approach requires users implementing an \emph{adapter} that provides target-specific code templates for each pattern type: \emph{match templates} that recognize a node and extract its components, and \emph{transform templates} that construct a node from bound variables. The adapter may also invoke backend-specific helper functions for operations such as predicate decomposition based on column dependencies, which many backends already provide as a utility.

For \calcite, the adapter generates a Java class extending \texttt{RelRule} for each \rulescript rule. The match phase emits an \texttt{operandSupplier} skeleton describing which plan shapes the rule applies to (e.g., \(\op{Filter}(\op{Filter}(\cdot))\) for filter merge). The transform phase emits the body of the \texttt{onMatch} callback, using \calcite's \texttt{RelBuilder} stack-based API to construct the rewritten plan from the matched fragments.

For \cockroach, \autoref{lst:optgen-example} shows the generated \optgen rule for \cockroach's optimizer corresponding to the \op{SemiJoin}-Aggregate transpose (\autoref{eq:semijoin-agg-rule}). The adapter maps \(\op{Aggregate}\) to \texttt{GroupBy} and \(\op{SemiJoin}\) to \texttt{SemiJoin}.

\begin{lstlisting}[caption={Generated \optgen rule for \op{SemiJoin}-Aggregate transpose.}, label={lst:optgen-example}]
[SemiJoinAggregateTranspose, Normalize]
(SemiJoin
    (GroupBy $input:* $aggs:*
        $private:(GroupingPrivate
            $groupingCols:* $ordering:*))
    $right:* $on:* &
        (OnlyRefsCols $on $groupingCols)
    $semiPrivate:*)
=>
(GroupBy
    (SemiJoin $input $right
        (RemapPredicate $on $groupingCols $input)
        $semiPrivate)
    $aggs $private)
\end{lstlisting}

The guard \texttt{OnlyRefsCols} ensures that the join predicate depends only on grouping columns, corresponding to the type constraint \(P : (K, Y) \to \op{Bool}\). The helper \texttt{RemapPredicate} rewrites the predicate from post-aggregation to pre-aggregation columns, corresponding to the substitution \(P(k,y) \to P(G(x),y)\). The generated code integrates with \cockroach's existing toolchain alongside manually implemented rules.

\section{Evaluation} \label{sec:eval}

We evaluate \rulescript along four research questions:
\begin{enumerate}
    \item[\textbf{RQ1.}] \emph{Coverage.} How expressive is \rulescript's core language and extension mechanism?
    \item[\textbf{RQ2.}] \emph{Portability.} What effort is required to support a new backend via \rulescript's adapter mechanism?
    \item[\textbf{RQ3.}] \emph{Effort.} How does \rulescript impact implementation complexity compared to native rule definitions?
    \item[\textbf{RQ4.}] \emph{Performance.} Do rules specified in \rulescript and automatically translated across backends yield measurable runtime improvements?
\end{enumerate}

We implemented \rulescript on top of the \qedp prover~\cite{wang2024qed} and integrated it with three backends via \rulescript's adapter mechanisms: \calcite (by generating Java code and integrating it with its optimizer), \cockroach (by generating \cockroach's \optgen DSL code and integrating it with its optimizer), and \datafusion (by writing an interpreter in Rust). Together these cover both execution modes described in \autoref{sec:execution} and three different real-world optimizer architectures. We assess runtime impact using the TPC-H benchmark.

\subsection{Rule Coverage} \label{sec:eval:coverage}

We selected 33 rules from \calcite's 91 core rewrite rules and encoded them in \rulescript. The selection covers all core relational operator types and all six transformation categories that are represented in \calcite (\autoref{tab:calcite-transformation-types}), prioritizing rules that exercise diverse pattern structures (e.g., single-operator rewrites, multi-operator transposes). The remaining 58 rules are excluded for two reasons: they involve operators not yet supported by \qedp (\op{Sort}, \op{Window}, \op{Sample}), or they are expression-level simplifications outside \rulescript's scope, such as constant folding in scalar expressions. Neither limitation reflects a restriction of \rulescript's rule language itself. All 33 rules were successfully verified by \qedp, with each proof completing within 5 seconds.

\autoref{tab:calcite-rules} shows the distribution of implemented rules across different operator categories. Again, the rules span all core relational operators, with particularly strong representation in filter and join categories, which together account for the majority of rewrites employed in modern optimizers.

\begin{table}[ht]
    \caption{Implemented rules by relational operator type. Rules involving multiple operators are counted in each relevant category.}
    \label{tab:calcite-rules}
    \begin{tabular}{lccccc} 
        \toprule
        & Aggregation & Filter & Project & Join & SetOp \\
        \midrule
        Implemented & 7 & 14 & 7 & 11 & 9 \\
        \bottomrule
    \end{tabular}
\end{table}

\autoref{tab:calcite-transformation-types} further categorizes the rules by their transformation pattern: \emph{Transpose} rules reorder adjacent operators; \emph{Merge} rules combine operators of the same type; \emph{Pushdown} rules move predicates closer to base relations; \emph{Join Transformations} include commutativity and join-specific rewrites; \emph{Simplification} rules eliminate unnecessary computations; and \emph{Expansion} rules transform plans into structurally different but equivalent forms. \rulescript achieves the highest coverage in merge (7/11) and pushdown (4/6) transformations, which are natural targets for algebraic pattern matching. The largest gap in terms of absolute numbers is in the simplification category (11/32), where the uncovered rules are predominantly expression-level simplifications involving specific numerical or Boolean operators (e.g., constant folding, predicate normalization), which fall outside of \rulescript's focus as they do not involve any query operator rewrites.
The remaining gaps in transpose, join transformations, expansion, and other categories are primarily due to operators not currently supported by the solver used by \rulescript, such as \op{Sort}, \op{Window}, and \op{Sample}.

\begin{table}[ht]
    \caption{Implemented rules by transformation pattern}
    \label{tab:calcite-transformation-types}
    \begin{tabular}{lcc} 
        \toprule
        Transformation Type & Implemented & Total \\ 
        \midrule
        Transpose & 6 & 21 \\
        Merge & 7 & 11 \\
        Pushdown & 4 & 6 \\
        Join Transformations & 3 & 8 \\
        Simplification & 11 & 32 \\
        Expansion & 2 & 7 \\
        Other & 0 & 6 \\
        \midrule
        \textbf{Total} & \textbf{33} & \textbf{91} \\
        \bottomrule
    \end{tabular}
\end{table}

We implemented the rules once in \rulescript, and then used \rulescript's adapter mechanisms to port 
all 33 rules to \calcite, \cockroach, and \datafusion, with the exception of set-operation rules in \datafusion: its internal representation of set operators differs from \calcite's, preventing direct translation of these rules. In addition, the \datafusion implementation includes rules that use the extension mechanism (\autoref{sec:extension}) to define \op{Left SemiJoin} and \op{Right SemiJoin} as custom operators, demonstrating that backend-specific operators can participate in verified rules alongside core operators.

The remaining gaps in rule coverage stem from limitations of the underlying solver, not of \rulescript's language. Operators such as \op{Sort} and \op{Window} require list semantics, whereas \qedp currently operates under bag semantics. \op{Sample} requires a probabilistic framework that \qedp does not support. These limitations are unrelated to \rulescript's ability to express rewrite rules, and they could be lifted by extending the solver in future work, as we are unaware of any solver that currently can handle such operators.

\subsection{Backend Portability} \label{sec:eval:portability}

Supporting a new backend in \rulescript requires implementing either a code-generation adapter or an interpreter, without any need to re-implement each rule. We evaluate this one-time integration cost across three architecturally different backends. \autoref{tab:adapter-loc} decomposes the implementation effort by the pipeline stages described in \autoref{sec:pipeline}: the \emph{match} phase (recognizing concrete plan nodes and extracting bindings), the \emph{transform} phase (constructing the rewritten plan from captured bindings), and any \emph{scaffold} code (context definitions, code emission templates or trait definitions, and utilities).

\begin{table}[ht]
    \caption{Backend integration effort (lines of code), decomposed by pipeline stage.}
    \label{tab:adapter-loc}
    \begin{tabular}{lccc}
        \toprule
        & \calcite & \cockroach & \datafusion \\
        \midrule
        Match    & 223 & 387 & 636 \\
        Transform & 237 & 296 & 743 \\
        Scaffold & 58  & 153 & 278 \\
        \midrule
        \textbf{Total} & \textbf{518} & \textbf{836} & \textbf{1,657} \\
        \bottomrule
    \end{tabular}
\end{table}

\paragraph{Code generation}
The \calcite and \cockroach adapters follow the code-generation path (\autoref{sec:code-generation}). The \calcite adapter emits Java classes that extend \calcite's \texttt{RelRule} framework~\cite{calcite2026relrule}: the match phase generates an \texttt{operandSupplier} skeleton describing which plan shapes the rule applies to, and the transform phase emits the body of the \texttt{onMatch} callback using \calcite's \texttt{RelBuilder} API to construct the rewritten plan. Because \calcite's framework handles plan traversal and rule scheduling, the adapter remains compact, with match and transform roughly balanced at around 230 lines each.

The \cockroach adapter targets \optgen~\cite{taft2020cockroachdb}, a domain-specific language for specifying optimizer transformations in CockroachDB. Each generated rule is an S-expression pairing a match pattern with a replace expression. The match phase notably requires more lines of code than \calcite's (387 vs.\ 223) because \optgen's pattern syntax encodes applicability constraints inline, such as checking column dependencies or join commutativity conditions, while the \calcite adapter instead delegates such checks to its \texttt{RelRule} framework. More code is needed for scaffolding as the adapter must emit the complete \optgen file structure and integrate with CockroachDB's built-in helper functions, as well as custom helpers for operations that lack native \optgen counterparts (e.g., predicate decomposition over uninterpreted symbols). A notable difference from \cockroach's native \optgen workflow is that \rulescript's adapter additionally supports uninterpreted symbols, which the standard \optgen language does not as it matches on concrete query operator types. This enables rules to remain abstract and reusable across contexts; the adapter handles the lowering into backend-specific constructs during code generation.

\paragraph{Interpreter}
The \datafusion adapter follows the interpreter path (\autoref{sec:interpreter}), operating directly on DataFusion's \texttt{LogicalPlan}~\cite{datafusion2026plan} representation at runtime. Unlike the code-generation adapters, the interpreter must implement the full matching and plan-construction logic itself rather than delegating to a target framework. This accounts for its larger size overall. The transform phase (743 lines) is the single largest component across all three backends because instantiating plans at runtime requires explicit column remapping and expression rewriting. Operations that code-generation adapters can delegate to backend utilities like \calcite's \texttt{RelBuilder} or \optgen's helper functions.

\paragraph{Discussion}
The two approaches offer different tradeoffs. Code-generation adapters are more compact because they leverage the target framework's pattern-matching and plan-construction infrastructure, but they require the target to provide such a framework. The interpreter approach takes more lines of code to implement but is self-contained: it can integrate with any backend that exposes a plan representation, without requiring a rule DSL or builder API. In both cases, the cost is paid once in building the adapter, as adding a new verified rule subsequently requires only writing the \rulescript definition, with no changes to the adapter at all.

The adapter and interpreter are not themselves formally verified: they are trusted components that translate verified rules into executable form. However, three properties limit the trust surface. First, the structured pipeline (\autoref{sec:pipeline}) constrains the translation to a systematic traversal of the rule's patterns, reducing the scope for errors compared to ad-hoc implementation. Second, the adapter is written once per backend and shared across all rules; a bug in the adapter would manifest across many rules, making it likely to be caught during testing. Third, the per-rule unit tests described in \autoref{sec:eval:validation} validate each rule's end-to-end behavior, catching adapter bugs that produce structurally incorrect rewrites.

To put these integration costs in context, we next compare the per-rule effort of \rulescript definitions against their native implementations.

\subsection{Implementation Effort} \label{sec:eval:effort}

We now compare the per-rule implementation effort of \rulescript definitions against native implementations in \calcite and \cockroach. We exclude \datafusion from this per-rule comparison because its native optimizer does not implement rules individually; instead, it bundles multiple transformations into monolithic passes, making a per-rule line count comparison infeasible (we discuss this structure at the end of this subsection). \autoref{tab:loc-comparison} groups the 33 rules by the transformation type categories from \autoref{tab:calcite-transformation-types} and reports total lines of code for each system.

\begin{table}[ht]
    \centering
    \caption{Lines of code by transformation category across three systems.}
    \label{tab:loc-comparison}
    \begin{tabular}{lcccc}
        \toprule
        Category & \rulescript & \calcite & \cockroach \\
        \midrule
        Transpose       &  129 & 1,173 &  119 \\
        Merge           &  155 &   787 &  122 \\
        Pushdown        &  123 &   301 &   71 \\
        Join Trans.     &   99 &   549 &   65 \\
        Simplification  &  335 &   804 &  155 \\
        Expansion       &  210 &   263 &   62 \\
        \midrule
        \textbf{Total}  & \textbf{1,051} & \textbf{3,877} & \textbf{594} \\
        \midrule
        Median per rule &   21 &    94 &   18 \\
        \bottomrule
    \end{tabular}
\end{table}

\rulescript definitions are approximately 3.7$\times$ more compact than their native \calcite implementations (with a median of 21 vs.\ 94 lines per rule). The gap is largest in the transpose rules (9$\times$), where \calcite requires substantial imperative logic for column-index manipulation when reordering operators. For example, \textsc{FilterProjectTranspose}, which pushes a filter below a projection, requires 274 lines in \calcite but only 20 in \rulescript. \calcite's native implementation must manually remap every column reference in the filter predicate from the projection's output indices to its input indices. In \rulescript, the composition \(P(f(x))\) expresses this remapping declaratively: the projection function \(f\) maps source columns to output columns, so composing \(P\) with \(f\) naturally rewrites the predicate in terms of the source schema. The gap is smallest in expansion rules (1.3$\times$), which transform plans into structurally more complex equivalent forms, like decomposing a single operator into multiple operators or introducing auxiliary constructs. In these cases, the specification must describe the new structure explicitly, so its size is dominated by the transformation's inherent complexity rather than by column bookkeeping or framework boilerplate that \rulescript can abstract away.

For a small number of rules, \rulescript definitions are comparable to or slightly longer than their \calcite counterparts. For instance, the \textsc{PruneEmptyFilter} rule requires 20 lines in \rulescript versus 16 in \calcite, and \textsc{JoinConditionPush} requires 55 versus 44. These cases arise because each \rulescript definition carries a fixed overhead for schema and symbol declarations that is proportionally larger for simple rules. \autoref{lst:prune-empty-filter} shows the \datafusion implementation of \textsc{PruneEmptyFilter} using \rulescript's Rust macro DSL: the core rewrite is a two-line match/transform pair, but the schema and predicate symbol declarations account for most of its size. The same rule exists in \rulescript's \calcite based implementation, but is slightly longer due to Java's more verbose syntax. In \calcite's native implementation, the simplest rules can be expressed as short imperative methods with minimal boilerplate, but this advantage disappears for more complex rules where the imperative logic grows substantially while the declarative specification does not.

\begin{lstlisting}[caption={\textsc{PruneEmptyFilter} in \rulescript's 
\datafusion macro-based DSL: filtering an empty relation yields empty.}, 
label={lst:prune-empty-filter}]
rule! {
    PruneEmptyFilterRule {
        schemas: { source: (col: T) },
        functions: { P(T) -> Bool },
        from: filter!(empty!(source), P(col)),
        to: empty!(source),
    }
}
\end{lstlisting}

Three \rulescript definitions are notably larger than the median: \textsc{ProjectAggregateMerge} (120 lines), \textsc{UnionPullUpConstants} (149 lines), and \textsc{UnionToDistinct} (61 lines). \textsc{ProjectAggregateMerge} eliminates unused aggregate calls: when a projection references only a subset of an aggregation's outputs, the unreferenced calls are dropped and the remaining column indices adjusted accordingly, requiring the rule to track which outputs survive and rewrite downstream references. The two union rules involve similar structural bookkeeping: \textsc{UnionPullUpConstants} must enumerate branches and factor out shared constant columns, while \textsc{UnionToDistinct} decomposes a duplicate-eliminating union into a union-all followed by a group-by-all deduplication. In each case, the size reflects genuine semantic complexity rather than representational overhead.

CockroachDB's \optgen rules are more compact still (median of 18 lines, total of 594 lines), which is expected: \optgen is a purpose-built pattern-rewrite DSL with concise match/replace syntax and no need for the type and schema declarations that \rulescript requires for verification. The comparison shows that \rulescript definitions are in the same order of magnitude as a dedicated DSL, while additionally providing formal verification and cross-backend portability that \optgen does not.

For \datafusion, the same rules are expressed in Rust using a macro-based DSL with a median of 20 lines per rule, comparable to the \rulescript Java definitions. This confirms that the specification cost is largely language-independent: the rule's semantic content, not the host language, determines its size.

In contrast, \datafusion's native optimizer rule implementation bundles the same transformations into monolithic passes. For example, its \texttt{push\_down\_filter} pass (1,435 lines excluding tests) implements filter merging, filter-projection transpose, filter-aggregate transpose, filter pushdown into joins, and filter movement past Sort, Window, and Unnest operators in a single tightly coupled traversal. Similarly, the \texttt{optimize\_projections} (1,163 lines) and \texttt{propagate\_empty\_relation} (232 lines) passes each combine multiple transformations into a single pass. This monolithic structure makes individual transformations difficult to verify, test, or reuse independently. \rulescript decomposes the same functionality into individually specified and verified units that can be ported across backends without modification.

\subsection{End-to-End Validation} \label{sec:eval:validation}

Rules verified by \qedp and translated by an adapter or interpreter must still integrate correctly with the target framework's optimizer pipeline: each rule must fire on applicable plans, produce structurally valid output, and participate in the optimizer's search strategy alongside native rules. We validate this at two levels: per-rule correctness testing and workload-level integration.

\paragraph{Per-rule correctness}
For \cockroach, we constructed test queries for every imported rule based on the \cockroach test files, following the same format used by \cockroach developers for \texttt{Optgen}. For each of the 33 rules, we confirmed that the corresponding \rulescript rule was triggered by the intended SQL queries and that the resulting optimized expression tree matched the expected semantics. For \datafusion, we ported 79 unit tests from \calcite's test suite, covering all 22 implemented rules. Each test constructs a \datafusion \texttt{LogicalPlan}, applies the rule through the interpreter, and asserts that the result matches the expected rewritten plan. All 79 tests passed.

\paragraph{Workload-level integration}
We ran the 22 standard TPC-H queries \cite{tpc-h} through both backends with the \rulescript-imported rules enabled. For \cockroach, we compared the vanilla optimizer against an extended configuration augmented with the 33 generated \texttt{Optgen} rules. Of the 22 queries, 18 triggered at least one imported rule during logical optimization, with several queries applying multiple rules iteratively before reaching a fixed point. The extended optimizer achieved a geometric mean speedup of 1.5$\times$ over the baseline; the improvement is driven primarily by filter pushdown and join reordering rules that reduce intermediate cardinalities in multi-join queries. For \datafusion, 14 of the 22 queries were optimized by at least one \rulescript rule through the interpreter.

Rule application overhead is not a concern in our prototype: the optimizer's search strategy and cost estimation dominate optimization time, and individual rule matching and transformation add no observable latency in either backend.

In both backends, rules written once in \rulescript and automatically translated participate in real optimization pipelines without manual per-rule intervention.

\section{Related Work}
\paragraph{Extensible query optimizers}
Rule-based query optimization originates with Starburst~\cite{pirahesh1992extensible}, whose design was generalized by the Volcano~\cite{graefe1993volcano} and Cascades~\cite{graefe1995cascades} frameworks. Modern systems build on this architecture but expose diverse rule interfaces. \calcite~\cite{begoli2018apache} defines rules as imperative Java classes that implement pattern matching and plan construction through the \texttt{RelRule} API. \cockroach~\cite{taft2020cockroachdb} uses \optgen, a declarative pattern-rewrite DSL that compiles to Go. \datafusion~\cite{datafusion2026project} implements rules as Rust functions that traverse and rewrite \texttt{LogicalPlan} trees in monolithic passes. \snowflake~\cite{dageville2016snowflake} similarly provide framework-specific rule mechanisms. In each case, rules are tightly coupled to the target framework's representation and API, requiring reimplementation when the same logical transformation is needed in a different system. \rulescript decouples rule specification from backend-specific execution by providing a shared language whose definitions can be compiled or interpreted for any supported backend.

\paragraph{Query equivalence and verified optimization}
Query equivalence is undecidable in general~\cite{trahtenbrot1963impossibility}, but restricted fragments admit automated verification. Early results cover conjunctive queries under bag semantics~\cite{ioannidis1995containment}. More recent solvers extend coverage to larger SQL fragments: \cosette~\cite{chu2017cosette} encodes equivalence as an SMT problem, \spes~\cite{zhou2022spes} uses symbolic reasoning, and \qedp~\cite{wang2024qed} reduces equivalence to algebraic reasoning over semiring expressions, delegating uninterpreted-symbol reasoning to the underlying SMT solver. \qedp currently offers the broadest coverage of SQL features among automated solvers and is the verifier used by \rulescript.

Applying formal verification to optimizer rules specifically has received less attention. \cosette has been used to check individual rewrite rules, but operates on concrete schemas rather than universally quantified patterns. HottSQL~\cite{chu2017hottsql} uses homotopy type theory to prove rewrite rules correct, but requires manual proof construction. \rulescript bridges this gap by encoding rules with uninterpreted symbols that \qedp can verify universally, producing proofs that hold for all schema instantiations without manual effort.

\section{Conclusion}
We presented \rulescript, a portable rule language for specifying query optimizer rewrite rules that are formally verified and executable across different database systems. Rules are written once using uninterpreted types, functions, and plans that abstract away schema-specific details, and an extension mechanism accommodates backend-specific operators such as semi-joins by defining their semantics in terms of core operators. \qedp verifies each rule for all possible instantiations of the uninterpreted symbols, providing correctness guarantees that testing alone cannot achieve. A shared execution pipeline supports both code generation (for backends with existing rule frameworks such as \calcite and \cockroach) and interpretation (for backends like \datafusion), with each backend requiring only a one-time adapter implementation. We evaluated \rulescript on 33 rules across three architecturally different backends; all rules were verified within seconds, and the translated rules fire on standard TPC-H queries in both \cockroach and \datafusion without manual per-rule intervention.

\balance
\clearpage

\bibliographystyle{ACM-Reference-Format}
\bibliography{cite}

@article{astrahan1976system,
  title={System R: Relational approach to database management},
  author={Astrahan, Morton M. and Blasgen, Mike W. and Chamberlin, Donald D. and Eswaran, Kapali P. and Gray, Jim N and Griffiths, Patricia P. and King, W Frank and Lorie, Raymond A. and McJones, Paul R. and Mehl, James W. and others},
  journal={ACM Transactions on Database Systems (TODS)},
  volume={1},
  number={2},
  pages={97--137},
  year={1976},
  publisher={ACM New York, NY, USA}
}

@inproceedings{begoli2018apache,
  title={Apache calcite: A foundational framework for optimized query processing over heterogeneous data sources},
  author={Begoli, Edmon and Camacho-Rodr{\'\i}guez, Jes{\'u}s and Hyde, Julian and Mior, Michael J and Lemire, Daniel},
  booktitle={Proceedings of the 2018 International Conference on Management of Data},
  pages={221--230},
  year={2018}
}

@article{chen1993combine,
  author={Chen, M.-S. and Yu, P.S.},
  journal={IEEE Transactions on Knowledge and Data Engineering}, 
  title={Combining joint and semi-join operations for distributed query processing}, 
  year={1993},
  volume={5},
  number={3},
  pages={534-542},
  doi={10.1109/69.224205}
}

@inproceedings{chu2017cosette,
  title={Cosette: An Automated Prover for SQL.},
  author={Chu, Shumo and Wang, Chenglong and Weitz, Konstantin and Cheung, Alvin},
  booktitle={CIDR},
  pages={1--7},
  year={2017}
}

@article{chu2017hottsql,
  title={HoTTSQL: Proving query rewrites with univalent SQL semantics},
  author={Chu, Shumo and Weitz, Konstantin and Cheung, Alvin and Suciu, Dan},
  journal={ACM SIGPLAN Notices},
  volume={52},
  number={6},
  pages={510--524},
  year={2017},
  publisher={ACM New York, NY, USA}
}

@inproceedings{dageville2016snowflake,
  title={The snowflake elastic data warehouse},
  author={Dageville, Benoit and Cruanes, Thierry and Zukowski, Marcin and Antonov, Vadim and Avanes, Artin and Bock, Jon and Claybaugh, Jonathan and Engovatov, Daniel and Hentschel, Martin and Huang, Jiansheng and others},
  booktitle={Proceedings of the 2016 International Conference on Management of Data},
  pages={215--226},
  year={2016}
}

@article{graefe1995cascades,
  title={The cascades framework for query optimization},
  author={Graefe, Goetz},
  journal={IEEE Data Eng. Bull.},
  volume={18},
  number={3},
  pages={19--29},
  year={1995}
}

@inproceedings{graefe1993volcano,
  title={The volcano optimizer generator: Extensibility and efficient search},
  author={Graefe, Goetz and McKenna, William J},
  booktitle={Proceedings of IEEE 9th international conference on data engineering},
  pages={209--218},
  year={1993},
  organization={IEEE}
}

@article{ioannidis1995containment,
  title={Containment of conjunctive queries: Beyond relations as sets},
  author={Ioannidis, Yannis E and Ramakrishnan, Raghu},
  journal={ACM Transactions on Database Systems (TODS)},
  volume={20},
  number={3},
  pages={288--324},
  year={1995},
  publisher={ACM New York, NY, USA}
}

@article{pirahesh1992extensible,
  title={Extensible/rule based query rewrite optimization in Starburst},
  author={Pirahesh, Hamid and Hellerstein, Joseph M and Hasan, Waqar},
  journal={ACM Sigmod Record},
  volume={21},
  number={2},
  pages={39--48},
  year={1992},
  publisher={ACM New York, NY, USA}
}

@inproceedings{rigger2020detect,
  author={Rigger, Manuel and Su, Zhendong},
  title={Detecting optimization bugs in database engines via non-optimizing reference engine construction},
  year={2020},
  isbn={9781450370431},
  publisher={Association for Computing Machinery},
  address={New York, NY, USA},
  url={https://doi.org/10.1145/3368089.3409710},
  doi={10.1145/3368089.3409710},
  pages={1140–1152},
  numpages={13},
  series={ESEC/FSE 2020}
}

@inproceedings{selinger1979access,
  title={Access path selection in a relational database management system},
  author={Selinger, P Griffiths and Astrahan, Morton M and Chamberlin, Donald D and Lorie, Raymond A and Price, Thomas G},
  booktitle={Proceedings of the 1979 ACM SIGMOD international conference on Management of data},
  pages={23--34},
  year={1979}
}

@inproceedings{taft2020cockroachdb,
  title={Cockroachdb: The resilient geo-distributed sql database},
  author={Taft, Rebecca and Sharif, Irfan and Matei, Andrei and VanBenschoten, Nathan and Lewis, Jordan and Grieger, Tobias and Niemi, Kai and Woods, Andy and Birzin, Anne and Poss, Raphael and others},
  booktitle={Proceedings of the 2020 ACM SIGMOD international conference on management of data},
  pages={1493--1509},
  year={2020}
}

@article{trahtenbrot1963impossibility,
  title={Impossibility of an algorithm for the decision problem in finite classes},
  author={Trahtenbrot, BA},
  journal={Nine Papers on Logic and Quantum Electrodynamics},
  pages={1--5},
  year={1963},
  publisher={American Mathematical Society}
}

@article{valduriez1984semijoin,
  author={Valduriez, Patrick and Gardarin, Georges},
  title={Join and Semijoin Algorithms for a Multiprocessor Database Machine},
  year={1984},
  issue_date={March 1984},
  publisher={Association for Computing Machinery},
  address={New York, NY, USA},
  volume={9},
  number={1},
  issn={0362-5915},
  url={https://doi.org/10.1145/348.318590},
  doi={10.1145/348.318590},
  journal={ACM Trans. Database Syst.},
  month=mar,
  pages={133–161},
  numpages={29}
}

@article{wang2024qed,
  title={QED: A Powerful Query Equivalence Decider for SQL},
  author={Wang, Shuxian and Pan, Sicheng and Cheung, Alvin},
  journal={Proc. VLDB Endow.},
  volume={17},
  number={11},
  pages={3602–3614},
  year={2024}
}

@inproceedings{zhou2022spes,
  title={SPES: A symbolic approach to proving query equivalence under bag semantics},
  author={Zhou, Qi and Arulraj, Joy and Navathe, Shamkant B and Harris, William and Wu, Jinpeng},
  booktitle={2022 IEEE 38th International Conference on Data Engineering (ICDE)},
  pages={2735--2748},
  year={2022},
  organization={IEEE}
}

@misc{calcite2026relrule,
  author={The Apache Calcite Team},
  title={Apache Calcite: RelRule},
  howpublished={\url{https://github.com/apache/calcite/blob/80e717ab2e5f9f6fb23995ab995264f2ec940882/core/src/main/java/org/apache/calcite/plan/RelRule.java}},
  year={2026},
  note={[Accessed 02-27-2026]},
}

@misc{cockroach2026optgen,
  author={The CockroachDB Team},
  title={CockroachDB: Optgen},
  howpublished={\url{https://github.com/cockroachdb/cockroach/tree/58e75c8e68cafc9a34482eb75f05ef51e6d8ca9a/pkg/sql/opt/optgen}},
  year={2026},
  note={[Accessed 02-27-2026]},
}

@misc{datafusion2026plan,
  author={The Apache Datafusion Team},
  title={Apache Datafusion: LogicalPlan},
  howpublished={\url{https://github.com/apache/datafusion/blob/6713439497561fa74a94177e5b8632322fb7cea5/datafusion/expr/src/logical_plan/plan.rs}},
  year={2026},
  note={[Accessed 02-27-2026]},
}

@misc{datafusion2026project,
  author={The Apache Datafusion Team},
  title={Apache DataFusion},
  howpublished={\url{https://datafusion.apache.org/}},
  year={2026},
  note={[Accessed 02-27-2026]},
}

@online{doris2022rule,
  author={Apache Doris},
  title={Transpose Semi-join Aggregation Rule},
  url={https://github.com/apache/doris/blob/8cb5a4210f794abca49b9aa0355120877aba2f32/fe/fe-core/src/main/java/org/apache/doris/nereids/rules/rewrite/TransposeSemiJoinAgg.java},
  year={2022},
}

@online{mysql2024bug,
  author={{MySQL}},
  title={{MySQL Bug \#114435: Incorrect query results caused by the subquery optimization}},
  url={https://bugs.mysql.com/bug.php?id=114435},
  year={2024}
}

@online{postgres2015bug,
  author={{PostgreSQL}},
  title={{PostgreSQL Bug \#13592: Optimizer throws out join constraint causing incorrect result}},
  url={https://www.postgresql.org/message-id/20150826195031.2091.40681%40wrigleys.postgresql.org},
  year={2015}
}

@online{spark2022rule,
  author={Apache Spark},
  title={Push Down Left Semi/Anti-join Rule},
  url={https://github.com/apache/spark/blob/168e5cf57147eed3e789b6d609a60396750094e4/sql/catalyst/src/main/scala/org/apache/spark/sql/catalyst/optimizer/PushDownLeftSemiAntiJoin.scala},
  year={2022},
}

@online{sqlite2025bug,
  author={{SQLite}},
  title={{Optimizer incorrectly push down aggregations to subqueries with distinct and union all.}},
  url={https://sqlite.org/forum/forumpost/a860f5fb2e},
  year={2025}
}

@misc{tpc-h,
  author={{Transaction Processing Performance Council (TPC)}},
  title={TPC-H Benchmark Specification, Version 2.17.1},
  howpublished={\url{https://www.tpc.org/tpc_documents_current_versions/pdf/tpc-h_v2.17.1.pdf}},
  year={2020},
  note={[Accessed 30-01-2026]}
}

\end{document}